\documentclass[aps,prl,twocolumn,longbibliography]{revtex4-2}

\usepackage{amssymb}
\usepackage{amsbsy}
\usepackage{amsmath}
\usepackage{graphicx}
\usepackage{graphics}
\usepackage{setspace}
\usepackage{array}
\usepackage{color}
\usepackage{fontenc}
\usepackage{textcomp}
\usepackage{bm}
\usepackage{float}
\usepackage[bookmarks=false,linkcolor=blue,urlcolor=blue,colorlinks,citecolor=blue]{hyperref}

\newcommand{\blue}[1]{{\color{blue}{#1}}}

\newcommand{\bsub}{\begin{subequations}}
\newcommand{\esub}{\end{subequations}}

\graphicspath{{../Figures/}}

\begin{document}
\title{Topological Reality Switch: Towards Bulk-Boundary Selective Lasing}
\author{Sayed Ali Akbar Ghorashi$^{1}$}\email[Correspondence\,to:\,]{sayedaliakbar.ghorashi@stonybrook.edu}
\author{Masatoshi Sato$^{2}$}
\affiliation{$^1$Department of Physics and Astronomy, Stony Brook University, Stony Brook, New York 11974, USA}
\affiliation{$^2$Yukawa Institute for Theoretical Physics, Kyoto University, Kyoto 606-8502, Japan}

\date{\today}

\newcommand{\be}{\begin{equation}}
\newcommand{\ee}{\end{equation}}
\newcommand{\bea}{\begin{eqnarray}}
\newcommand{\eea}{\end{eqnarray}}
\newcommand{\h}{\hspace{0.30 cm}}
\newcommand{\vs}{\vspace{0.30 cm}}
\newcommand{\n}{\nonumber}

\begin{abstract}
The emergence of complex spectra in non-Hermitian systems causes dramatic changes even under weak perturbations, significantly hindering their precise control for study and integration into practical applications. Achieving a controlled method to generate a real spectrum in non-Hermitian systems has long been a key objective in the field. In this study, we explore the 2D non-Hermitian Su-Schrieffer-Heeger (SSH) model and introduce a \emph{reality switch} that allows for the controllable induction of a real spectrum depending on the imposed boundary condition. We show that a topological phase transition in the complex gap accompanies the switching process. Our work lays the cornerstone for developing a selective bulk-boundary control mechanism for the gain and lasing behaviors in non-Hermitian systems. 
\end{abstract}
\maketitle

\blue{\emph{Introduction}}.---Non-Hermitian systems have received significant attention in recent years due to their intriguing physical properties and unique behaviors that deviate from conventional Hermitian systems. In contrast to Hermitian systems, which have real eigenvalues corresponding to observable quantities in quantum mechanics, non-Hermitian systems often feature complex spectra that can lead to phenomena such as exceptional points and the non-Hermitian skin effects \cite{reviewNHUeda,ReviewNHRMP,heiss2012physics,okuma2023non}.
Due to their high sensitivity to even weak perturbations, achieving control over the complex nature of the spectrum has long been a goal in the field. Heuristically, it has been shown that the presence of specific symmetries ensures the reality of the spectrum, with the combination of parity and time-reversal ($\mathcal{PT}$) symmetry being the most prominent example \cite{PhysRevLett.80.5243,mostafazadeh2002pseudo,PhysRevResearch.2.033391}. A few studies have taken a bottom-up approach to induce reality, although they still have some limitations in controlability \cite{PhysRevB.105.L100102,yang2022designing}.

\begin{figure}
    \centering
    \includegraphics[width=1\linewidth]{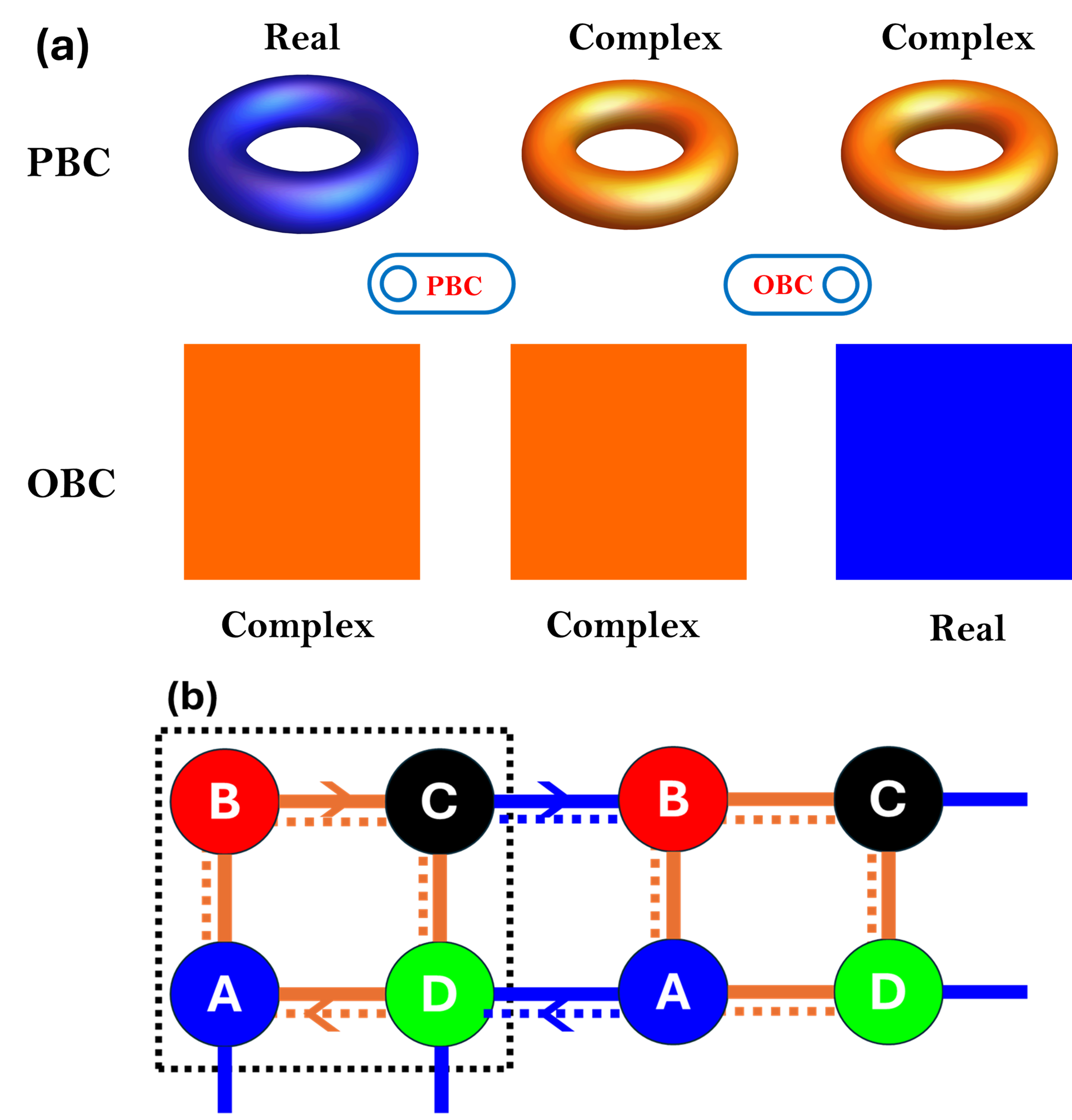}
    \caption{Schematics showing topological reality switch where a non-Hermitian Hamiltonian with complex spectrum (a) can selectively possess real spectrum in the presence of PBC or OBC. (b) lattice model of a 2D non-Hermitian SSH model, where the dotted square shows the unit cell. The orange and blue lines show intra- and inter-unit cell hoppings, respectively. Dashed lines denote the asymmetric hopping in different directions.}
    \label{fig:adpic}
\end{figure}

In this work, we study the 2D non-Hermitian Su-Schriffer-Heeger (SSH) model and discover a perturbation that acts as a \emph{reality switch} that can turn on/off the complex spectrum in  a selective and controllable way; see Fig.~\ref{fig:adpic}. We provide a mechanism that leads to the reality of the spectrum for either open or periodic boundary conditions. Furthermore, we show that switching spectrum complexity between bulk and boundary is accompanied by a topological phase transition. Finally, we discuss possible generalizations of the proposed reality switch.

\begin{figure*}[tb!]
    \centering
    \includegraphics[width=1\linewidth]{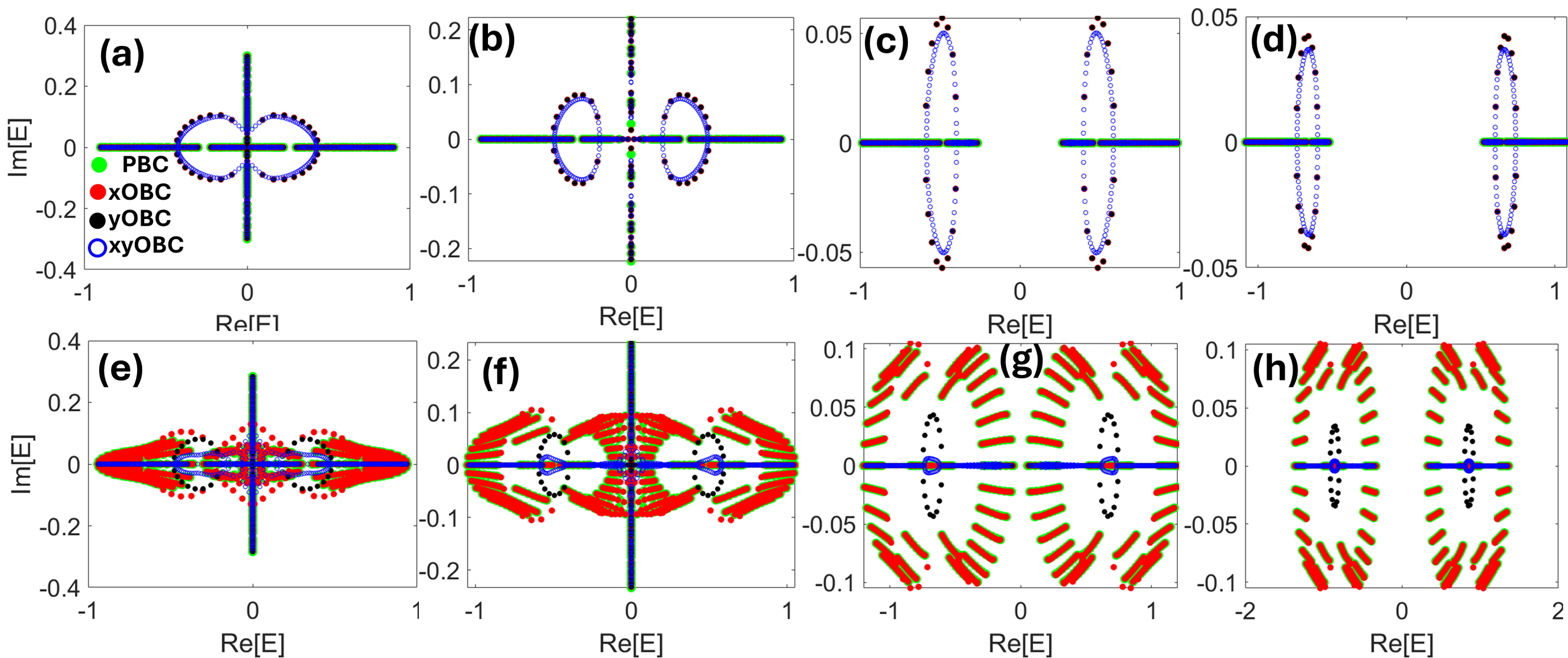}
    \caption{Evolution of complex gap for $\gamma_{in}=0.2,\,\gamma_{ex}=0.4,\,\gamma'_{in,ex}=0.5\gamma_{in,ex}$ and (a) $\alpha=0,\beta=0$, (b) $\alpha=0.2,\beta=0$ (c) $\alpha=0.4,\beta=0$, (d) $\alpha=0.6,\beta=0$, (e) $\alpha=0,\beta=0.2$ (f) $\alpha=0,\beta=0.4$ (g) $\alpha=0,\beta=0.6$ (h) $\alpha=0,\beta=0.8$. Green, red, black, and blue represent PBC, $x$OBC, $y$OBC, and full OBC ($xy$OBC), deceptively, as is shown in the inset of (a). }
    \label{fig:IZ_ZI_separate}
\end{figure*}
\blue{\emph{Model}}.---We start with the 2D non-Hermitian SSH model in Fig.\ref{fig:adpic}(d), where non-Hermiticity is induced via asymmetric hoppings $\gamma_{\rm in, ex}$, $\gamma'_{\rm in, ex}\in \mathbb{R}$ \cite{Tangprx}. The corresponding matrix Hamiltonian reads
\begin{align}\label{2dssh}
    &H^{\rm NH}_{\rm SSH}({\bm k},\gamma_{\rm in,ex})= h({\bm k},\gamma_{\rm in,ex}) + h^{\dagger}({\bm k},\gamma'_{\rm in,ex}); \cr
    & h({\bm k},\gamma_{\rm in,ex}) =\left[
  \begin{array}{cc|cc}
    0 & \gamma_{\rm in} & 0 & \gamma_{\rm ex}e^{-ik_x} \\
     \gamma_{\rm ex}e^{ik_y} & 0 & \gamma_{\rm in} & 0 \\
     \hline
    0 &  \gamma_{\rm ex}e^{ik_x} & 0 & \gamma_{\rm in} \\
    \gamma_{\rm in} & 0 &  \gamma_{\rm ex}e^{-ik_y} & 0 \\
  \end{array}
\right], \cr
\end{align}
where the $(i,j)$-component gives the hopping between $i$ and $j$ internal states $(i,j=A,B,C,D)$ in the unit cell in Fig.\ref{fig:adpic}(b).
The symmetric limit $\gamma_{\rm in,ex}=\gamma'_{\rm in,ex}$ reduces to the Hermitian 2D SSH.
As one can see immediately in Fig.\ref{fig:adpic} (b),
this model has several symmetries; inversion (${\cal P}$), $\mathcal{P} H({\bm k}) \mathcal{P}^{\dagger}= H(-{\bm k})$, time-reversal (${\cal T}$), $\mathcal{T} H({\bm k}) \mathcal{T}^{\dagger}= H^*(-{\bm k})$, combinations of reciprocal and mirror reflections (${\cal RM}_{x,y}$), $\mathcal{R}\mathcal{M}_{x,y} H(k_x,k_y) (\mathcal{R}\mathcal{M}_{x,y})^{\dagger}= H^T(\pm k_x,\mp k_y)$, four-fold rotation (${\cal C}_4$), $\mathcal{C}_4H(k_x,k_y)\mathcal{C}_4^\dagger=H(k_y,-k_x)$, and sublattice (${\cal S}$), $\mathcal{S} H({\bm k}) \mathcal{S}^{\dagger}= -H({\bm k})$, where $\mathcal{P}=\sigma^x\sigma^0$, $\mathcal{T}= \sigma^0\sigma^0$, $\mathcal{R}\mathcal{M}_{x(y)}=\sigma^x\sigma^x(\sigma^0\sigma^x)$, ${\cal C}_4=\sigma^0(\sigma^x-i\sigma^y)/2+\sigma^x(\sigma^x+i\sigma^y)/2$ and $\mathcal{S}=\sigma^0\sigma^z$. Here, $\sigma^\mu$ is the Pauli matrix, and $\sigma^\mu\sigma^\nu$ is the Kronecker product, of which $4\times 4$ components correspond to transitions between four sites in the unit cell.
In the following, we investigate the effect of two types of on-site potentials (= diagonal components of $\sigma^\mu\sigma^\nu$) in which they can selectively induce reality in the periodic boundary condition (PBC) or open boundary condition (OBC) spectrum.

\blue{\emph{PBC reality switch}}.--- We begin with the on-site potential $\alpha \sigma^0\sigma^z$ ($\alpha\in\mathbb{R}$).
This potential preserves ${\cal P}$, ${\cal T}$, ${\cal RM}_{x,y}{\cal S}$,
and ${\cal RM}_{x,y}{\cal C}_4$ symmetries.
As shown immediately,
a sufficiently large $\alpha$ induces the real bulk spectrum under PBC, ensured by ${\cal P}{\cal T}$ symmetry.

Figure \ref{fig:IZ_ZI_separate}(a) shows the spectrum of Eq.~\eqref{2dssh} for various boundary conditions with $\gamma_{in}=0.2,\,\gamma_{ex}=0.4,\,\gamma'_{in,ex}=0.5\gamma_{in,ex}$ and $\alpha=0$. Under full PBC, the spectrum consists of four branches: Two with purely real energy and two forming a crossing of purely imaginary and purely real energies.
The former branches are in the ${\cal P}{\cal T}$ symmetric phase, where the $\mathcal{PT}$ symmetry ensures the reality of the spectrum.
Meanwhile, the latter are partially in the ${\cal P}{\cal T}$ broken phase; thus, their spectra are complex.
For $\alpha \neq 0$, $\mathcal{PT}$ is still preserved, and, as such, the reality of the ${\cal P}{\cal T}$ symmetric branches. When increasing $\alpha$ (Figs.\ref{fig:IZ_ZI_separate} (b)-(d)), the ${\cal P}{\cal T}$ broken branches show the ${\cal P}{\cal T}$ phase transition, then the whole branches exhibit real spectra, as shown in Figs.~\ref{fig:IZ_ZI_separate}(c) and (d).
These bulk spectra display arcs, not areas, in the complex energy plane, thereby preserving reality without non-Hermitian skin effects even under OBCs.

Under OBCs, the system also hosts edge modes.
The edge modes have complex spectra even in the ${\cal P}{\cal T}$ symmetric phase, because the boundary explicitly breaks $\mathcal{PT}$ symmetry.
Notably, like the bulk modes, the complex modes also do not show the non-Hermitian skin effect:
Whereas these complex modes under the open boundary condition in the $x$-direction ($x$OBC) (and the $y$-direction ($y$OBC)) show loops in the spectrum, they have opposite winding numbers on opposite boundaries due to ${\cal P}{\cal T}$ symmetry.
Furthermore, the edge modes on opposite boundaries of $x$OBC ($y$OBC) can easily mix under the full OBC via the edge modes under $y$OBC ($x$OBC).
($x$OBC and $y$OBC support edge modes simultaneously because of ${\cal R}{\cal M}_{x,y}{\cal C}_4$ symmetry.)
Thus, the spectral winding numbers and, consequently, the corresponding skin effects are canceled.

\blue{\emph{OBC reality switch}}.----Next, we study another on-site potential
 $\beta \sigma^z\sigma^0$ ($\beta\in \mathbb{R})$.
 See Figs.\ref{fig:IZ_ZI_separate} (e)-(h).
 The presence of $\beta$ keeps symmetries
 $\mathcal{RM}_y$, $\mathcal{T}$, and ${\cal RM}_x{\cal S}$.
 As shown below, the former two allow for real spectra under full OBC, using the non-Hermitian skin effects.

First, ${\cal R}{\cal M}_y$ symmetry, ${\cal R}{\cal M}_y H(k_x,k_y)({\cal R}{\cal M}_y)^\dagger
=H^T(-k_x,k_y)$, ensures that there is no spectral winding number in the $x$-direction, so there is no skin effect under $x$OBC.
Thus, the bulk spectra under $x$OBC are identical to those under PBC: The difference in the spectra between $x$OBC and PBC originates from the edge modes under $x$OBC.
As shown in Figs. \ref{fig:IZ_ZI_separate} (g)-(h),
edge modes under $x$OBC disappear for sufficiently large $\beta$.

The non-Hermitian skin effect occurs under $y$OBC.
Remarkably, for large $\beta$ (Figs. \ref{fig:IZ_ZI_separate} (g)-(h)),
the non-Hermitian skin effect results in the real bulk spectrum.
We find that ${\cal R}{\cal M}_y{\cal T}$ symmetry
ensures the reality of the bulk spectrum:
Since there is no non-Hermitian skin effect in the $x$-direction, the non-Bloch theory in the $y$-direction provides the bulk spectrum under $y$OBC (and thus, that under full OBC without the non-Hermitian skin effect in the $x$-direction.):
For each $k_x$, one can introduce the non-Bloch Hamiltonian $H(k_x,\beta_y)$ by replacing $e^{ik_y}$ in $H(k_x,k_y)$ with $\beta_y$ \cite{YM2023}, then, the modular condition \cite{YM2019, SS1960, Delvaux2012} for $|\beta_y|$ determines the bulk spectrum.
When the non-Herminian skin effect occurs, the spectrum for each $k_x$ forms arcs in the complex energy plane.
Notably, the arcs can realize the ${\cal R}{\cal M}_y{\cal T}$ symmetry in a manner different from that under PBC:
If they are in the non-Bloch ${\cal R}{\cal M}_y{\cal T}$ symmetric phase,
which is an analog of the non-Bloch ${\cal P}{\cal T}$ symmetric phase \cite{Longhi19-1,Longhi19-2,SWW2022}, the arcs are invariant under ${\cal R}{\cal M}_y{\cal T}$, and thus their energies become real.

While the edge modes under $y$OBC have complex energies, they also display (almost) a real spectrum under full OBC, where the deviation from the real energy vanishes in the thermodynamic limit.
${\cal T}$ symmetry is responsible for this property:
Although the loop spectra of the edge modes under $y$OBC host opposite winding numbers on opposite boundaries due to ${\cal R}{\cal M}_y{\cal T}$ symmetry, they rarely mix under the full OBC because no additional edge mode appears on the boundaries normal to the $x$-direction.
Thus, the loops shrink under full OBC due to the non-Hermitian skin effect.
In particular, in the thermodynamic limit, the mixing vanishes, so the loops become arcs with real spectra in the non-Bloch ${\cal T}$ symmetric phase.
Therefore, for large $\beta$, the entire spectrum is real under full OBC.

\begin{figure}[tb!]
    \centering
    \includegraphics[width=1.\linewidth]{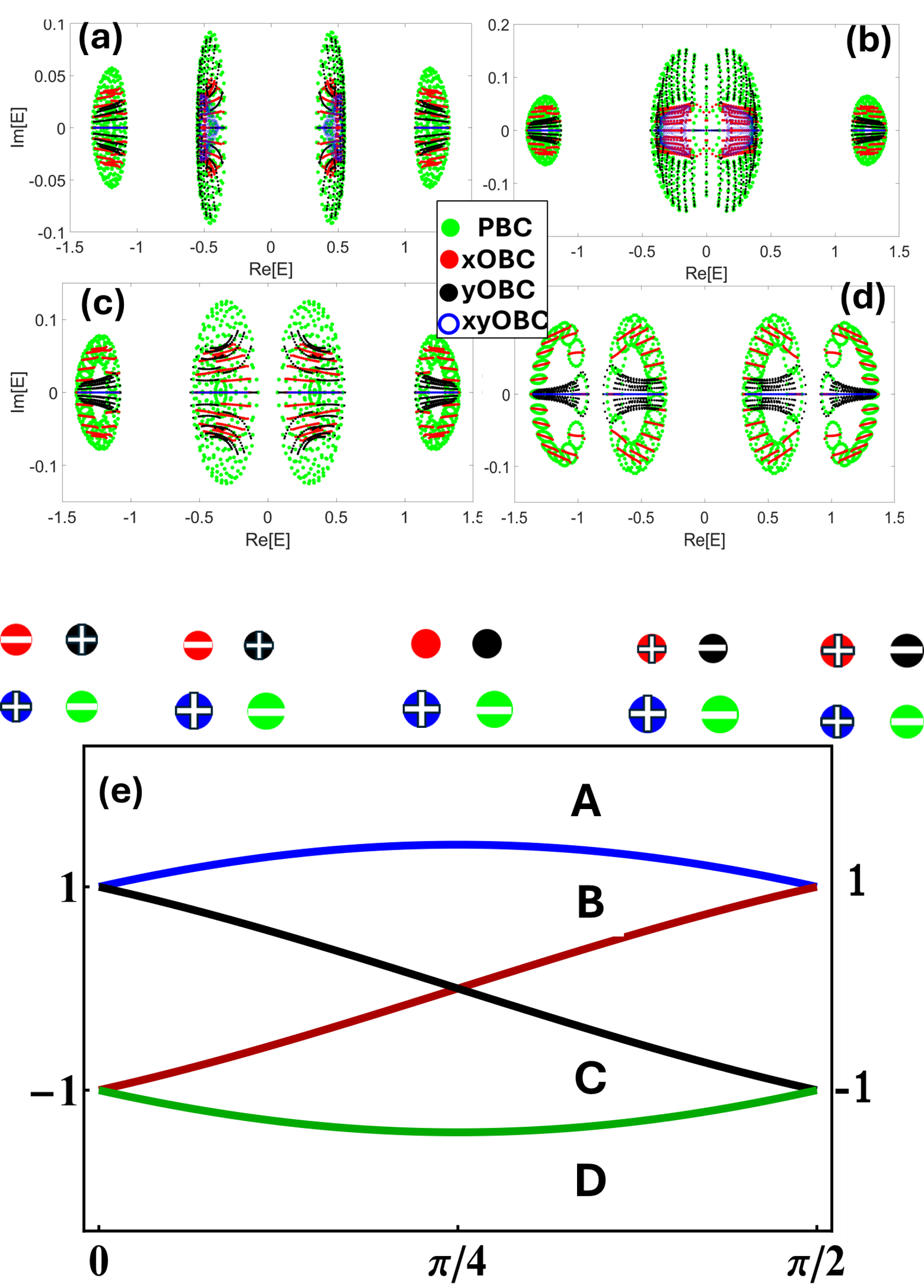}
    \caption{The complex gap plot for \eqref{2dssh} in presence of perturbation $0.8\times (\cos(\theta) \sigma^0\sigma^3 + \sin(\theta) \sigma^3\sigma^0)$ with (a) $\theta=0.15 \pi$ (b) $\theta=0.25 \pi$, (c) $\theta=0.35 \pi$, (d) $\theta=0.45 \pi$. }
    \label{fig:costhIZ_sinthZI}
\end{figure}

\blue{\emph{On-demand PBC-OBC reality switch}}.----So far, we have examined two types of potentials; one induces reality in the spectrum under PBC, and the other does so under full OBC. Next, we demonstrate the evolution of the two potentials, illustrating how their combination functions as a reality switch that selectively enforces the real spectrum based on the boundary conditions. Figure \ref{fig:costhIZ_sinthZI} shows the evolution of the complex spectrum for the combined potential $\cos\theta\sigma^0\sigma^3+\sin\theta\sigma^3\sigma^0$ ($0<\theta<\pi/2$).
As is evident, as soon as one goes away from $\theta=0$ (Fig.~\ref{fig:costhIZ_sinthZI}(a)), $\mathcal{PT}$ is broken, and the entire spectra, including under PBC, now becomes complex. They remain complex until $\theta=0.25\pi$, where the line-gap between two bands in the center closes. Interestingly, as $\theta$ increases, the line-gap reopens, and the spectrum under full OBC becomes real.
Importantly, for $\theta\neq \pi/2$, the $\mathcal{RM}_y{\cal T}$ is now explicitly broken, and thus it can not ensure the reality of the OBC spectrum.
Still, ${\cal T}$ symmetry remains for the combined potential, which suppresses the imaginary part of the OBC spectrum if the system size is small enough \cite{SWW2022}.


To gain a deeper understanding of the microscopic behavior of these potentials, Fig.~\ref{fig:costhIZ_sinthZI}(e) illustrates how each component of the combined potentials changes with $\theta$. In particular, the topological transition at $\theta=\pi/4$ is associated with a redistribution of the signs of the on-site potential across different internal states. Starting from $\theta=0$, the on-site potential forms a balanced quadrupole that is inversion symmetric. As $\theta$ deviates from zero, an imbalance is generated between internal states where $A, D$ have larger magnitudes compared to $B, C$, which signals broken $\mathcal{PT}$. Exactly at $\theta=\pi/4$ the magnitude of on-site potential vanishes on sites $B, C$. Upon reopening the lin-gap away from $\theta=\pi/4$ the potential sign on $B$ and $C$ swaps, leading to the formation of a pair of dipoles. Moreover, considering that ${\rm Im} E$ of spectrum under full OBC is much more suppressed for $\pi/4<\theta<\pi/2$ by comparing Figs.~\ref{fig:costhIZ_sinthZI}(c,d) and Figs.~\ref{fig:IZ_ZI_separate}(g,h) (Im$E =\mathcal{O}(10^{-3})$ in Fig.~\ref{fig:IZ_ZI_separate}(h)), we conclude that (i) the topological phase transition corresponds to a point where the potential is turned off on a pair of site, and (ii) an imbalanced dipole potential is favored for inducing reality under full OBC.  \\
With this intuition, we can now propose another on-site potential that satisfies these conditions. For example, the behavior of the potential of the form $0.8\times (\cos(\theta) \sigma^0\sigma^3 + \sin(\theta) \sigma^3\sigma^3)$ is the same as in the case in Fig.~\ref{fig:costhIZ_sinthZI} except that at the critical point of $\theta=\pi/4$ potential vanishes on the $C$ and $D$ sites.

\blue{\emph{Concluding remarks}}.---We close with a few remarks. First, in this work, we specifically focused on the on-site potentials; however, it is noteworthy that this is not limited to the on-site potentials. For example, we have found inter-site perturbations that can work as a reality switch for bulk and boundary, e.g., $\sigma^1\sigma^1+\sigma^2\sigma^2$. Moreover, we found a combination of perturbations that can induce reality for all boundary conditions simultaneously, e.g., $\sigma^1\sigma^0+\sigma^1\sigma^3$. We leave a detailed investigation of these perturbations for future work.
Second, while here we have focused mainly on a specific parameter regime of the 2D SSH model, the reality switch described here persists in the 2D Hatano-Nelson limit, $\gamma_{in}=\gamma_{ex},\,\gamma'_{in}=\gamma'_{ex}$. A future research direction would be to generalize the results here to 3D systems and investigate possible implications for higher-order topology and boundaries \cite{GLHHOWSM,NHHHODSM,NHHOWSM}.
Finally, many variations of non-Hermitian SSH models have been realized on different platforms \cite{zou2021observation, PhysRevResearch.6.023140} with immense tunability. Therefore, we expect that the topological reality switch proposed here can be immediately realized in various setups.

\emph{Acknowledgment}.---Authors gratefully acknowledge support from the Simons Center for Geometry and Physics, Stony Brook University, at which some or all of the research for this paper was performed. M.S. is supported by JSPS KAKENHI (Grants Nos. JP24K00569 and JP25H01250) and JST CREST (Grants No. JPMJCR19T2).

\bibliography{main.bib}

\begin{thebibliography}{22}%
\makeatletter
\providecommand \@ifxundefined [1]{%
 \@ifx{#1\undefined}
}%
\providecommand \@ifnum [1]{%
 \ifnum #1\expandafter \@firstoftwo
 \else \expandafter \@secondoftwo
 \fi
}%
\providecommand \@ifx [1]{%
 \ifx #1\expandafter \@firstoftwo
 \else \expandafter \@secondoftwo
 \fi
}%
\providecommand \natexlab [1]{#1}%
\providecommand \enquote  [1]{``#1''}%
\providecommand \bibnamefont  [1]{#1}%
\providecommand \bibfnamefont [1]{#1}%
\providecommand \citenamefont [1]{#1}%
\providecommand \href@noop [0]{\@secondoftwo}%
\providecommand \href [0]{\begingroup \@sanitize@url \@href}%
\providecommand \@href[1]{\@@startlink{#1}\@@href}%
\providecommand \@@href[1]{\endgroup#1\@@endlink}%
\providecommand \@sanitize@url [0]{\catcode `\\12\catcode `\$12\catcode
  `\&12\catcode `\#12\catcode `\^12\catcode `\_12\catcode `\%12\relax}%
\providecommand \@@startlink[1]{}%
\providecommand \@@endlink[0]{}%
\providecommand \url  [0]{\begingroup\@sanitize@url \@url }%
\providecommand \@url [1]{\endgroup\@href {#1}{\urlprefix }}%
\providecommand \urlprefix  [0]{URL }%
\providecommand \Eprint [0]{\href }%
\providecommand \doibase [0]{https://doi.org/}%
\providecommand \selectlanguage [0]{\@gobble}%
\providecommand \bibinfo  [0]{\@secondoftwo}%
\providecommand \bibfield  [0]{\@secondoftwo}%
\providecommand \translation [1]{[#1]}%
\providecommand \BibitemOpen [0]{}%
\providecommand \bibitemStop [0]{}%
\providecommand \bibitemNoStop [0]{.\EOS\space}%
\providecommand \EOS [0]{\spacefactor3000\relax}%
\providecommand \BibitemShut  [1]{\csname bibitem#1\endcsname}%
\let\auto@bib@innerbib\@empty
\bibitem [{\citenamefont {Ashida}\ \emph {et~al.}(2020)\citenamefont {Ashida},
  \citenamefont {Gong},\ and\ \citenamefont {Ueda}}]{reviewNHUeda}%
  \BibitemOpen
  \bibfield  {author} {\bibinfo {author} {\bibfnamefont {Y.}~\bibnamefont
  {Ashida}}, \bibinfo {author} {\bibfnamefont {Z.}~\bibnamefont {Gong}},\ and\
  \bibinfo {author} {\bibfnamefont {M.}~\bibnamefont {Ueda}},\ }\bibfield
  {title} {\bibinfo {title} {Non-hermitian physics},\ }\href
  {https://doi.org/10.1080/00018732.2021.1876991} {\bibfield  {journal}
  {\bibinfo  {journal} {Advances in Physics}\ }\textbf {\bibinfo {volume}
  {69}},\ \bibinfo {pages} {249–435} (\bibinfo {year} {2020})}\BibitemShut
  {NoStop}%
\bibitem [{\citenamefont {Bergholtz}\ \emph {et~al.}(2021)\citenamefont
  {Bergholtz}, \citenamefont {Budich},\ and\ \citenamefont
  {Kunst}}]{ReviewNHRMP}%
  \BibitemOpen
  \bibfield  {author} {\bibinfo {author} {\bibfnamefont {E.~J.}\ \bibnamefont
  {Bergholtz}}, \bibinfo {author} {\bibfnamefont {J.~C.}\ \bibnamefont
  {Budich}},\ and\ \bibinfo {author} {\bibfnamefont {F.~K.}\ \bibnamefont
  {Kunst}},\ }\bibfield  {title} {\bibinfo {title} {Exceptional topology of
  non-hermitian systems},\ }\href
  {https://doi.org/10.1103/RevModPhys.93.015005} {\bibfield  {journal}
  {\bibinfo  {journal} {Rev. Mod. Phys.}\ }\textbf {\bibinfo {volume} {93}},\
  \bibinfo {pages} {015005} (\bibinfo {year} {2021})}\BibitemShut {NoStop}%
\bibitem [{\citenamefont {Heiss}(2012)}]{heiss2012physics}%
  \BibitemOpen
  \bibfield  {author} {\bibinfo {author} {\bibfnamefont {W.}~\bibnamefont
  {Heiss}},\ }\bibfield  {title} {\bibinfo {title} {The physics of exceptional
  points},\ }\href@noop {} {\bibfield  {journal} {\bibinfo  {journal} {Journal
  of Physics A: Mathematical and Theoretical}\ }\textbf {\bibinfo {volume}
  {45}},\ \bibinfo {pages} {444016} (\bibinfo {year} {2012})}\BibitemShut
  {NoStop}%
\bibitem [{\citenamefont {Okuma}\ and\ \citenamefont
  {Sato}(2023)}]{okuma2023non}%
  \BibitemOpen
  \bibfield  {author} {\bibinfo {author} {\bibfnamefont {N.}~\bibnamefont
  {Okuma}}\ and\ \bibinfo {author} {\bibfnamefont {M.}~\bibnamefont {Sato}},\
  }\bibfield  {title} {\bibinfo {title} {Non-hermitian topological phenomena: A
  review},\ }\href@noop {} {\bibfield  {journal} {\bibinfo  {journal} {Annual
  Review of Condensed Matter Physics}\ }\textbf {\bibinfo {volume} {14}},\
  \bibinfo {pages} {83} (\bibinfo {year} {2023})}\BibitemShut {NoStop}%
\bibitem [{\citenamefont {Bender}\ and\ \citenamefont
  {Boettcher}(1998)}]{PhysRevLett.80.5243}%
  \BibitemOpen
  \bibfield  {author} {\bibinfo {author} {\bibfnamefont {C.~M.}\ \bibnamefont
  {Bender}}\ and\ \bibinfo {author} {\bibfnamefont {S.}~\bibnamefont
  {Boettcher}},\ }\bibfield  {title} {\bibinfo {title} {Real spectra in
  non-hermitian hamiltonians having $\mathcal{P}\mathcal{T}$ symmetry},\ }\href
  {https://doi.org/10.1103/PhysRevLett.80.5243} {\bibfield  {journal} {\bibinfo
   {journal} {Phys. Rev. Lett.}\ }\textbf {\bibinfo {volume} {80}},\ \bibinfo
  {pages} {5243} (\bibinfo {year} {1998})}\BibitemShut {NoStop}%
\bibitem [{\citenamefont {Mostafazadeh}(2002)}]{mostafazadeh2002pseudo}%
  \BibitemOpen
  \bibfield  {author} {\bibinfo {author} {\bibfnamefont {A.}~\bibnamefont
  {Mostafazadeh}},\ }\bibfield  {title} {\bibinfo {title} {Pseudo-hermiticity
  versus pt symmetry: The necessary condition for the reality of the spectrum
  of a non-hermitian hamiltonian},\ }\href@noop {} {\bibfield  {journal}
  {\bibinfo  {journal} {Journal of Mathematical Physics}\ }\textbf {\bibinfo
  {volume} {43}},\ \bibinfo {pages} {205} (\bibinfo {year} {2002})}\BibitemShut
  {NoStop}%
\bibitem [{\citenamefont {Kawabata}\ and\ \citenamefont
  {Sato}(2020)}]{PhysRevResearch.2.033391}%
  \BibitemOpen
  \bibfield  {author} {\bibinfo {author} {\bibfnamefont {K.}~\bibnamefont
  {Kawabata}}\ and\ \bibinfo {author} {\bibfnamefont {M.}~\bibnamefont
  {Sato}},\ }\bibfield  {title} {\bibinfo {title} {Real spectra in
  non-hermitian topological insulators},\ }\href
  {https://doi.org/10.1103/PhysRevResearch.2.033391} {\bibfield  {journal}
  {\bibinfo  {journal} {Phys. Rev. Res.}\ }\textbf {\bibinfo {volume} {2}},\
  \bibinfo {pages} {033391} (\bibinfo {year} {2020})}\BibitemShut {NoStop}%
\bibitem [{\citenamefont {Long}\ \emph {et~al.}(2022)\citenamefont {Long},
  \citenamefont {Xue},\ and\ \citenamefont {Zhang}}]{PhysRevB.105.L100102}%
  \BibitemOpen
  \bibfield  {author} {\bibinfo {author} {\bibfnamefont {Y.}~\bibnamefont
  {Long}}, \bibinfo {author} {\bibfnamefont {H.}~\bibnamefont {Xue}},\ and\
  \bibinfo {author} {\bibfnamefont {B.}~\bibnamefont {Zhang}},\ }\bibfield
  {title} {\bibinfo {title} {Non-hermitian topological systems with eigenvalues
  that are always real},\ }\href {https://doi.org/10.1103/PhysRevB.105.L100102}
  {\bibfield  {journal} {\bibinfo  {journal} {Phys. Rev. B}\ }\textbf {\bibinfo
  {volume} {105}},\ \bibinfo {pages} {L100102} (\bibinfo {year}
  {2022})}\BibitemShut {NoStop}%
\bibitem [{\citenamefont {Yang}\ \emph {et~al.}(2022)\citenamefont {Yang},
  \citenamefont {Tan}, \citenamefont {Tai}, \citenamefont {Koh}, \citenamefont
  {Li}, \citenamefont {Longhi},\ and\ \citenamefont {Lee}}]{yang2022designing}%
  \BibitemOpen
  \bibfield  {author} {\bibinfo {author} {\bibfnamefont {R.}~\bibnamefont
  {Yang}}, \bibinfo {author} {\bibfnamefont {J.~W.}\ \bibnamefont {Tan}},
  \bibinfo {author} {\bibfnamefont {T.}~\bibnamefont {Tai}}, \bibinfo {author}
  {\bibfnamefont {J.~M.}\ \bibnamefont {Koh}}, \bibinfo {author} {\bibfnamefont
  {L.}~\bibnamefont {Li}}, \bibinfo {author} {\bibfnamefont {S.}~\bibnamefont
  {Longhi}},\ and\ \bibinfo {author} {\bibfnamefont {C.~H.}\ \bibnamefont
  {Lee}},\ }\bibfield  {title} {\bibinfo {title} {Designing non-hermitian real
  spectra through electrostatics},\ }\href@noop {} {\bibfield  {journal}
  {\bibinfo  {journal} {Science Bulletin}\ }\textbf {\bibinfo {volume} {67}},\
  \bibinfo {pages} {1865} (\bibinfo {year} {2022})}\BibitemShut {NoStop}%
\bibitem [{\citenamefont {Tang}\ \emph {et~al.}(2021)\citenamefont {Tang},
  \citenamefont {Agudo-Canalejo},\ and\ \citenamefont {Golestanian}}]{Tangprx}%
  \BibitemOpen
  \bibfield  {author} {\bibinfo {author} {\bibfnamefont {E.}~\bibnamefont
  {Tang}}, \bibinfo {author} {\bibfnamefont {J.}~\bibnamefont
  {Agudo-Canalejo}},\ and\ \bibinfo {author} {\bibfnamefont {R.}~\bibnamefont
  {Golestanian}},\ }\bibfield  {title} {\bibinfo {title} {Topology protects
  chiral edge currents in stochastic systems},\ }\href
  {https://doi.org/10.1103/PhysRevX.11.031015} {\bibfield  {journal} {\bibinfo
  {journal} {Phys. Rev. X}\ }\textbf {\bibinfo {volume} {11}},\ \bibinfo
  {pages} {031015} (\bibinfo {year} {2021})}\BibitemShut {NoStop}%
\bibitem [{\citenamefont {Yokomizo}\ and\ \citenamefont
  {Murakami}(2023)}]{YM2023}%
  \BibitemOpen
  \bibfield  {author} {\bibinfo {author} {\bibfnamefont {K.}~\bibnamefont
  {Yokomizo}}\ and\ \bibinfo {author} {\bibfnamefont {S.}~\bibnamefont
  {Murakami}},\ }\bibfield  {title} {\bibinfo {title} {Non-bloch bands in
  two-dimensional non-hermitian systems},\ }\href
  {https://doi.org/10.1103/PhysRevB.107.195112} {\bibfield  {journal} {\bibinfo
   {journal} {Phys. Rev. B}\ }\textbf {\bibinfo {volume} {107}},\ \bibinfo
  {pages} {195112} (\bibinfo {year} {2023})}\BibitemShut {NoStop}%
\bibitem [{\citenamefont {Yokomizo}\ and\ \citenamefont
  {Murakami}(2019)}]{YM2019}%
  \BibitemOpen
  \bibfield  {author} {\bibinfo {author} {\bibfnamefont {K.}~\bibnamefont
  {Yokomizo}}\ and\ \bibinfo {author} {\bibfnamefont {S.}~\bibnamefont
  {Murakami}},\ }\bibfield  {title} {\bibinfo {title} {Non-bloch band theory of
  non-hermitian systems},\ }\href
  {https://doi.org/10.1103/PhysRevLett.123.066404} {\bibfield  {journal}
  {\bibinfo  {journal} {Phys. Rev. Lett.}\ }\textbf {\bibinfo {volume} {123}},\
  \bibinfo {pages} {066404} (\bibinfo {year} {2019})}\BibitemShut {NoStop}%
\bibitem [{\citenamefont {Schmidt}\ and\ \citenamefont
  {Spitzer}(1960)}]{SS1960}%
  \BibitemOpen
  \bibfield  {author} {\bibinfo {author} {\bibfnamefont {P.}~\bibnamefont
  {Schmidt}}\ and\ \bibinfo {author} {\bibfnamefont {F.}~\bibnamefont
  {Spitzer}},\ }\bibfield  {title} {\bibinfo {title} {The toeplitz matrices of
  an arbitrary laurent polynomial},\ }\href@noop {} {\bibfield  {journal}
  {\bibinfo  {journal} {Math. Scand.}\ }\textbf {\bibinfo {volume} {8}},\
  \bibinfo {pages} {15} (\bibinfo {year} {1960})}\BibitemShut {NoStop}%
\bibitem [{\citenamefont {Delvaux}(2012)}]{Delvaux2012}%
  \BibitemOpen
  \bibfield  {author} {\bibinfo {author} {\bibfnamefont {S.}~\bibnamefont
  {Delvaux}},\ }\bibfield  {title} {\bibinfo {title} {Equilibrium problem for
  the eigenvalues of banded block toeplitz matrices},\ }\href@noop {}
  {\bibfield  {journal} {\bibinfo  {journal} {Math. Nachr.}\ }\textbf {\bibinfo
  {volume} {285}},\ \bibinfo {pages} {1935} (\bibinfo {year}
  {2012})}\BibitemShut {NoStop}%
\bibitem [{\citenamefont {Longhi}(2019{\natexlab{a}})}]{Longhi19-1}%
  \BibitemOpen
  \bibfield  {author} {\bibinfo {author} {\bibfnamefont {S.}~\bibnamefont
  {Longhi}},\ }\bibfield  {title} {\bibinfo {title} {Probing non-hermitian skin
  effect and non-bloch phase transitions},\ }\href
  {https://doi.org/10.1103/PhysRevResearch.1.023013} {\bibfield  {journal}
  {\bibinfo  {journal} {Phys. Rev. Res.}\ }\textbf {\bibinfo {volume} {1}},\
  \bibinfo {pages} {023013} (\bibinfo {year} {2019}{\natexlab{a}})}\BibitemShut
  {NoStop}%
\bibitem [{\citenamefont {Longhi}(2019{\natexlab{b}})}]{Longhi19-2}%
  \BibitemOpen
  \bibfield  {author} {\bibinfo {author} {\bibfnamefont {S.}~\bibnamefont
  {Longhi}},\ }\bibfield  {title} {\bibinfo {title} {Non-bloch pt symmetry
  breaking in non-hermitian photonic quantum walks},\ }\href
  {https://doi.org/10.1364/OL.44.005804} {\bibfield  {journal} {\bibinfo
  {journal} {Opt. Lett.}\ }\textbf {\bibinfo {volume} {44}},\ \bibinfo {pages}
  {5804} (\bibinfo {year} {2019}{\natexlab{b}})}\BibitemShut {NoStop}%
\bibitem [{\citenamefont {Song}\ \emph {et~al.}(2022)\citenamefont {Song},
  \citenamefont {Wang},\ and\ \citenamefont {Wang}}]{SWW2022}%
  \BibitemOpen
  \bibfield  {author} {\bibinfo {author} {\bibfnamefont {F.}~\bibnamefont
  {Song}}, \bibinfo {author} {\bibfnamefont {H.-Y.}\ \bibnamefont {Wang}},\
  and\ \bibinfo {author} {\bibfnamefont {Z.}~\bibnamefont {Wang}},\ }\bibinfo
  {title} {Non-bloch pt symmetry: Universal threshold and dimensional
  surprise},\ in\ \href {https://doi.org/10.1142/9789811264153_0017} {\emph
  {\bibinfo {booktitle} {A Festschrift in Honor of the C N Yang Centenary}}}\
  (\bibinfo  {publisher} {world scientific},\ \bibinfo {year} {2022})\ pp.\
  \bibinfo {pages} {299--311}\BibitemShut {NoStop}%
\bibitem [{\citenamefont {Ghorashi}\ \emph {et~al.}(2020)\citenamefont
  {Ghorashi}, \citenamefont {Li},\ and\ \citenamefont {Hughes}}]{GLHHOWSM}%
  \BibitemOpen
  \bibfield  {author} {\bibinfo {author} {\bibfnamefont {S.~A.~A.}\
  \bibnamefont {Ghorashi}}, \bibinfo {author} {\bibfnamefont {T.}~\bibnamefont
  {Li}},\ and\ \bibinfo {author} {\bibfnamefont {T.~L.}\ \bibnamefont
  {Hughes}},\ }\bibfield  {title} {\bibinfo {title} {Higher-order weyl
  semimetals},\ }\href {https://doi.org/10.1103/PhysRevLett.125.266804}
  {\bibfield  {journal} {\bibinfo  {journal} {Phys. Rev. Lett.}\ }\textbf
  {\bibinfo {volume} {125}},\ \bibinfo {pages} {266804} (\bibinfo {year}
  {2020})}\BibitemShut {NoStop}%
\bibitem [{\citenamefont {Ghorashi}\ \emph
  {et~al.}(2021{\natexlab{a}})\citenamefont {Ghorashi}, \citenamefont {Li},
  \citenamefont {Sato},\ and\ \citenamefont {Hughes}}]{NHHHODSM}%
  \BibitemOpen
  \bibfield  {author} {\bibinfo {author} {\bibfnamefont {S.~A.~A.}\
  \bibnamefont {Ghorashi}}, \bibinfo {author} {\bibfnamefont {T.}~\bibnamefont
  {Li}}, \bibinfo {author} {\bibfnamefont {M.}~\bibnamefont {Sato}},\ and\
  \bibinfo {author} {\bibfnamefont {T.~L.}\ \bibnamefont {Hughes}},\ }\bibfield
   {title} {\bibinfo {title} {Non-hermitian higher-order dirac semimetals},\
  }\href {https://doi.org/10.1103/PhysRevB.104.L161116} {\bibfield  {journal}
  {\bibinfo  {journal} {Phys. Rev. B}\ }\textbf {\bibinfo {volume} {104}},\
  \bibinfo {pages} {L161116} (\bibinfo {year}
  {2021}{\natexlab{a}})}\BibitemShut {NoStop}%
\bibitem [{\citenamefont {Ghorashi}\ \emph
  {et~al.}(2021{\natexlab{b}})\citenamefont {Ghorashi}, \citenamefont {Li},\
  and\ \citenamefont {Sato}}]{NHHOWSM}%
  \BibitemOpen
  \bibfield  {author} {\bibinfo {author} {\bibfnamefont {S.~A.~A.}\
  \bibnamefont {Ghorashi}}, \bibinfo {author} {\bibfnamefont {T.}~\bibnamefont
  {Li}},\ and\ \bibinfo {author} {\bibfnamefont {M.}~\bibnamefont {Sato}},\
  }\bibfield  {title} {\bibinfo {title} {Non-hermitian higher-order weyl
  semimetals},\ }\href {https://doi.org/10.1103/PhysRevB.104.L161117}
  {\bibfield  {journal} {\bibinfo  {journal} {Phys. Rev. B}\ }\textbf {\bibinfo
  {volume} {104}},\ \bibinfo {pages} {L161117} (\bibinfo {year}
  {2021}{\natexlab{b}})}\BibitemShut {NoStop}%
\bibitem [{\citenamefont {Zou}\ \emph {et~al.}(2021)\citenamefont {Zou},
  \citenamefont {Chen}, \citenamefont {He}, \citenamefont {Bao}, \citenamefont
  {Lee}, \citenamefont {Sun},\ and\ \citenamefont
  {Zhang}}]{zou2021observation}%
  \BibitemOpen
  \bibfield  {author} {\bibinfo {author} {\bibfnamefont {D.}~\bibnamefont
  {Zou}}, \bibinfo {author} {\bibfnamefont {T.}~\bibnamefont {Chen}}, \bibinfo
  {author} {\bibfnamefont {W.}~\bibnamefont {He}}, \bibinfo {author}
  {\bibfnamefont {J.}~\bibnamefont {Bao}}, \bibinfo {author} {\bibfnamefont
  {C.~H.}\ \bibnamefont {Lee}}, \bibinfo {author} {\bibfnamefont
  {H.}~\bibnamefont {Sun}},\ and\ \bibinfo {author} {\bibfnamefont
  {X.}~\bibnamefont {Zhang}},\ }\bibfield  {title} {\bibinfo {title}
  {Observation of hybrid higher-order skin-topological effect in non-hermitian
  topolectrical circuits},\ }\href
  {https://www.nature.com/articles/s41467-021-26414-5} {\bibfield  {journal}
  {\bibinfo  {journal} {Nature Communications}\ }\textbf {\bibinfo {volume}
  {12}},\ \bibinfo {pages} {7201} (\bibinfo {year} {2021})}\BibitemShut
  {NoStop}%
\bibitem [{\citenamefont {Slootman}\ \emph {et~al.}(2024)\citenamefont
  {Slootman}, \citenamefont {Cherifi}, \citenamefont {Eek}, \citenamefont
  {Arouca}, \citenamefont {Bergholtz}, \citenamefont {Bourennane},\ and\
  \citenamefont {Smith}}]{PhysRevResearch.6.023140}%
  \BibitemOpen
  \bibfield  {author} {\bibinfo {author} {\bibfnamefont {E.}~\bibnamefont
  {Slootman}}, \bibinfo {author} {\bibfnamefont {W.}~\bibnamefont {Cherifi}},
  \bibinfo {author} {\bibfnamefont {L.}~\bibnamefont {Eek}}, \bibinfo {author}
  {\bibfnamefont {R.}~\bibnamefont {Arouca}}, \bibinfo {author} {\bibfnamefont
  {E.~J.}\ \bibnamefont {Bergholtz}}, \bibinfo {author} {\bibfnamefont
  {M.}~\bibnamefont {Bourennane}},\ and\ \bibinfo {author} {\bibfnamefont
  {C.~M.}\ \bibnamefont {Smith}},\ }\bibfield  {title} {\bibinfo {title}
  {Breaking and resurgence of symmetry in the non-hermitian
  su-schrieffer-heeger model in photonic waveguides},\ }\href
  {https://doi.org/10.1103/PhysRevResearch.6.023140} {\bibfield  {journal}
  {\bibinfo  {journal} {Phys. Rev. Res.}\ }\textbf {\bibinfo {volume} {6}},\
  \bibinfo {pages} {023140} (\bibinfo {year} {2024})}\BibitemShut {NoStop}%
\end{thebibliography}%

\end{document}